\documentclass[runningheads,a4paper]{llncs}
\usepackage{makeidx}  
\usepackage{orcidlink}
\usepackage{multirow}
\usepackage[margin=1in]{geometry}
\usepackage{array}
\usepackage{tabularx}
\usepackage{array}
\usepackage{multirow}

\usepackage{amssymb}
\usepackage{graphicx}
\usepackage{subcaption}
\usepackage{float}
\usepackage{lscape}
\usepackage {xcolor}
\usepackage{hyperref}
\usepackage{comment}
\usepackage{makecell}
\usepackage{orcidlink}
\usepackage{xcolor}
\usepackage{booktabs}

\begin{document}
\mainmatter              
\title{AI and Agile Software Development: \newline A Research Roadmap from the XP2025 Workshop}


%
\titlerunning{Research Roadmap on AI and Agile}  
%


\author{Zheying Zhang\orcidlink{0000-0002-6205-4210}\inst{1}
\and Tomas Herda\orcidlink{0009-0005-2912-380X}\inst{2}
\and Victoria Pichler\orcidlink{0009-0006-4406-127X}\inst{3}
\and Pekka Abrahamsson\orcidlink{0000−0002−4360−2226}\inst{1}
\and Geir K. Hanssen\orcidlink{0000-0003-2718-6637}\inst{4}
\and Joshua Kerievsky\inst{5}
\and Alex Polyakov\inst{6}
\and Mohit Chandna\inst{7}
\and Marius Irgens\orcidlink{0009-0009-6784-1529}\inst{8}
\and Kai-Kristian Kemell\orcidlink{0000-0002-0225-4560}\inst{1}
\and Ayman Asad Khan\orcidlink{0009-0004-0134-8313}\inst{1}
\and Crystal Kwok\orcidlink{0009-0002-7818-0526}\inst{9}
\and Evan Leybourn\inst{10}
\and Munish Malik\inst{7}
\and Dorota Mleczko\inst{11}
\and Morteza Moalagh\orcidlink{0000-0002-7489-4583}\inst{12} 
\and Christopher Morales\inst{10}
\and Yuliia Pieskova \inst{13}
\and Daniel Planötscher\orcidlink{0009-0001-9162-0939} \inst{9}
\and Mika Saari\orcidlink{0000-0001-7677-2355}\inst{1}
\and Anastasiia Tkalich\orcidlink{0000-0001-7391-4194}\inst{4}
\and Karl Josef Gstettner\inst{14}
\and Xiaofeng Wang\orcidlink{0000-0001-8424-419X}\inst{9}
}
\institute{
\textit{Tampere University, Tampere, Finland
\and
AI Center of Excellence - Austrian Post, Vienna, Austria
\and
Digital Logistics Platform - Group-IT - Austrian Post, Vienna, Austria
\and
SINTEF, 034 Trondheim, Norway
\and
Industrial Logic, Inc., California, USA
\and
Project Simple, California, USA
\and 
Equal Experts, India
\and
University of Oslo, Oslo, Norway
\and
Free University of Bozen-Bolzano, Bolzano, Italy
\and
Business Agility Institute, California, USA
\and
Paloma's AI Academy, Genoa, Italy
\and
Norwegian University of Science and Technology, Trondheim, Norway
\and
Alpha Affinity, Munich, Germany
\and
AMS Austria, Vienna, Austria}
}
\authorrunning{Zhang, Herda, et al.}
\maketitle              
\begin{abstract}        
This paper synthesizes the key findings from a full-day XP2025 workshop on "AI and Agile: From Frustration to Success", held in Brugg-Windisch, Switzerland \cite{herda2025ai}. The workshop brought together over 30 interdisciplinary academic researchers and industry practitioners to tackle the concrete challenges and emerging opportunities at the intersection of Generative Artificial Intelligence (GenAI) and agile software development. Through structured, interactive breakout sessions, participants identified shared pain points like tool fragmentation, governance, data quality, and critical skills gaps in AI literacy and prompt engineering. These issues were further analyzed, revealing underlying causes and cross-cutting concerns. The workshop concluded by collaboratively co-creating a multi-thematic research roadmap, articulating both short-term, implementable actions and visionary, long-term research directions. This cohesive agenda aims to guide future investigation and drive the responsible, human-centered integration of GenAI into agile practices.
\end{abstract}

\section{Introduction and Evolution of AI and Agile}
\subsection{Introducing the XP2025 Workshop}

This paper presents a research roadmap synthesized from the "AI and Agile Software Development: From Frustration to Success" workshop, a full-day event held at the XP2025 conference in Brugg-Windisch, Switzerland \cite{herda2025ai,XP2025workshop_AI&Agile}. As the third consecutive AI and Agile event at XP, the workshop attracted \textbf{over 30 participants} from research and practice who shared a common interest in integrating artificial intelligence into agile ways of working. While a growing body of literature explores this synergy \cite{cinkusz2024towards,bahi2024integrating,zhang2024llm,manish2024autonomous,cabrero2024exploring,sami2024early,ji2023survey,kulkarni2017integration,liu2023pre,shneiderman2022human,nguyen2023generative,amershi2019guidelines}, a significant gap persists between theoretical possibilities and the practical experiences of agile teams, whose insights are essential for guiding effective research. The organizers set four objectives: to investigate the adoption of AI in Agile practices; to facilitate the exchange of practical experiences, including both challenges and successes; to develop a forward-looking research roadmap collaboratively; and to demonstrate the application of AI tools by integrating them into the workshop discussion. 

A competitive call for contributions resulted in \textbf{17 submissions}, from which two keynotes, three research papers, and three industrial experience talks were accepted for presentation. The committee itself formed an academia-industry coalition, reflecting the collaborative theme of the workshop.

A notable feature of the workshop was its deliberate use of AI-powered tools to enhance and document the proceedings. For instance, an AI-generated theme song was created to capture the event's collaborative spirit, and a custom GPT was developed and populated with all accepted papers, keynotes, and session outputs to serve as a persistent and interactive knowledge base. These resources\footnote{\scriptsize Final Workshop Program: \href{https://conf.researchr.org/home/xp-2025/aiandagile-2025\#program}{https://conf.researchr.org/home/xp-2025/aiandagile-2025} \newline Official Workshop Website: \href{https://gpt-lab.eu/ai-agile-workshop-xp2025/}{gpt-lab.eu} \newline Custom Conference GPT: \href{https://chatgpt.com/g/g-68341f9d6c8c8191847064ae11c08098-aiandagilexp2025gpt}{chatgpt.com} \newline AI-Generated Workshop Song: \href{https://suno.com/song/3123f552-9b00-486b-a552-87ad01bbbdac}{suno.com}\newline Workshop Opening Talk: \url{https://youtu.be/xigzwCzttV4}\newline Workshop Highlights Video: \url{https://www.youtube.com/watch?v=TmQwxuRAOIk}}, along with video recordings and the official program and workshop website, provide a comprehensive and lasting record of the discussions. The workshop thus served not only as a forum for discussing AI's role in agile but also as a live demonstration of its potential.

\subsection{Evolution of Agile and the Emergence of AI-powered Tooling}
Since its formalization in the \textbf{Agile Manifesto (2001)} \cite{Agilemanifesto}, the agile movement fundamentally reshaped the software engineering landscape. Its principles prioritized individuals and interactions, customer collaboration, and the ability to respond to change over rigid processes and comprehensive documentation. Agile frameworks such as Scrum, eXtreme Programming (XP), and methods like Kanban became industry standards, fostering iterative development, continuous feedback, and a focus on delivering value. For nearly two decades, the evolution in this space was driven by process refinement and human-centric collaboration practices. Early technical enablers such as \textbf{continuous integration (2000s)} \cite{Cruisecontrol} and \textbf{DevOps pipelines (2010s)} \cite{Briefhistorydevops} automated build, test, and deployment loops, reinforcing agility through faster feedback.

Concurrently, AI techniques matured along three overlapping waves:
\begin{itemize}
    \item \textbf{2005-2015 – Predictive analytics and classic ML} \cite{anvik2006bugfix,zimmermann2007defects}: Early defect-prediction models and recommendation engines appeared in issue tracking and backlog grooming tools.
    \item \textbf{2016-2020 – Deep learning for code and text}: Sequence-to-sequence networks powered code-completion, test-generation, and backlog-item classification. The launch of \textbf{GitHub Copilot (2021)}\cite{Githubcopilot} and \textbf{ChatGPT (late 2022)} marked a paradigm shift. Suddenly, AI was no longer a peripheral analyst but an active, conversational partner directly embedded in the developer's workflow signaling a tipping point in mainstream developer exposure to large language models (LLMs).
    \item \textbf{2021-present – Foundation models and multi-agent orchestration}: Modern foundation and reasoning models can generate user stories, draft pull-request descriptions, and critique each other’s outputs. These capabilities highlight the growing sophistication of AI-generated content, while underscoring the continued importance of keeping humans in the loop to ensure relevance, accuracy, and alignment with human intent. Both \textbf{Joshua Kerievsky}\cite{kerievsky2025xp} and \textbf{Alex Polyakov}\cite{polyakov2025xp}, keynote speakers at XP2025, explored how Agile practices can evolve in the age of GenAI. Kerievsky argued that the core practices of XP - such as pair programming, test-driven development, and collective ownership - provide the essential structure and quality control needed to steer and validate AI-generated outputs, turning a potentially erratic tool into a reliable collaborator. Similarly, Polyakov emphasized that AI tooling is most effective when designed not merely to automate existing frameworks but to address deeper team challenges around communication, clarity, and decision-making. This, he noted, can foster the right behaviours to enhance adaptability. Together, these insights frame a central challenge for the community: how to thoughtfully integrate the immense power of GenAI into the proven, human-centered philosophies of agile development.
\end{itemize}

\subsection{Motivation for this Research Roadmap}

This paper builds on the discussion initiated during the XP2024 Industry and Practice Track on AI and Agile\footnote{\scriptsize XP2024\href{https://agilealliance.org/xp2024/industry-and-practice-ai-and-agile/}{AI and Agile - Industry and Practice Track}}. While the 2024 track presented experiences from early adopters, the XP2025 workshop exposed broader. The identified community-wide research gaps emphasize the need for a structured research response. 

Three gaps dominated the roadmap session. They are
\begin{itemize}
    \item \textbf{Fragmented tooling}: “Too many tools, unclear which to use” received the highest number of votes during the frustration prioritization exercise. Teams face overlapping feature sets, inconsistent user interfaces and proprietary model endpoints that hinder end-to-end flow.
    \item \textbf{Governance and privacy uncertainty}: Participants expressed concern about data-protection boundaries and intellectual-property ownership when AI models ingest sprint artifacts. These anxieties delay or derail organizational adoption.
    \item \textbf{Skills and literacy gaps}: A pervasive lack of prompting know-how, together with early abandonment after poor first trials, leaves many teams unable to translate AI potential into value.
\end{itemize}

These gaps jeopardize core agile principles of sustainable pace, transparency and customer-driven value. Fragmented tooling disrupts flow and increases cognitive load, undermining a sustainable pace. Governance and privacy uncertainties reduce transparency in how AI models are applied and create barriers to open team collaboration. Skills and literacy gaps prevent effective use of AI to deliver incremental value aligned with customer needs. At the same time, the intense interest and rapidly evolving landscape argue for timely, coordinated research. By synthesizing workshop insights, historical evolution, and prior XP experience, the present roadmap seeks to move the community from isolated experiments to \textbf{evidence-based, human-centered integration of AI into agile software development}.

\section{Challenges in AI-Enhanced Agile}

\subsection{Promise and Paradox of AI in Agile}

The last three years have seen an explosion of large-language-model (LLM)–powered assistants, code copilots and natural-language agents that promise faster delivery cycles, higher quality, and richer insights across the agile life-cycle, from backlog refinement to production monitoring. Industry studies by GitHub 
\cite{github2023octoverse} and McKinsey 
\cite{mckinsey2023genai} report productivity gains of 20–50\% for common development tasks such as unit-test generation and boiler-plate coding. Academic evidence likewise shows that AI-driven tools can reduce cognitive load and foster knowledge sharing in distributed teams\cite{weisz2025chi,ulfsnes2024springer} and accelerate exploratory spikes through rapid prototyping \cite{subramonyam2025chi}. Yet, as underscored by the XP2025 workshop, translating this potential into sustained value remains fraught with practical obstacles. For instance, recent studies also point to emerging drawbacks, such as reduced collaboration and team interaction when developers overly rely on GenAI tools. Ulfsnes et al.\cite{ulfsnes2024springer} highlight this tension in their analysis of agile teams, and Wivestad et al.\cite{wivestad2024copilot} further explore how increased individual satisfaction with AI tools may introduce potential drawbacks for teams who rely on GenAI, such as reduced collaboration and weakened team cohesion.

\subsection{Synthesis of Practitioner Frustrations}

During the workshop’s retrospective session, the practitioners clustered more than 120 collected pain points into six overarching categories. The Padlet-based voting exercise \footnote{\scriptsize https://padlet.com/embed/hsxlg0sr4k1235rd} then prioritized the most pressing challenges within each category. The voting results from workshop participants are reported in the workshop summary \cite{herda2025ai}. Table \ref{tab:frustration-categories} summarizes the categories, their top-voted concerns and the share of votes received.  These data underpin the research roadmap that follows in Section 3.

\vspace{-15pt}

\begin{table}[htbp]
  \centering
  \caption{Frustration Categories, Most-Voted Challenges, \% of Votes, and Number of Voters}
  \label{tab:frustration-categories}
  \begin{tabular}{@{}clp{6.8cm}cc@{}}
    \toprule
    \textbf{Cat.} & \textbf{Frustration Category} & \textbf{Most-Voted Challenge} & \textbf{\% of Votes} & \textbf{Voters (n)} \\ \midrule
    F1 & \textbf{Tooling Challenges}           & Too many tools, unclear which to use              & 73.3\,\% & 15 \\
    F2 & \textbf{Governance \& Compliance}     & Unclear data privacy and protection boundaries    & 53.3\,\% & 15 \\
    F3 & \textbf{Team \& Process Misalignment} & AI integration doesn’t yield valuable outcomes    & 52.9\,\% & 17 \\
    F4 & \textbf{Data \& Model Quality Issues}        & Hallucinations and unreliable outputs             & 66.7\,\% & 18 \\
    F5 & \textbf{Knowledge \& Prompting Gaps}  & Lack of prompting skills or best practices        & 78.6\,\% & 14 \\
    F6 & \textbf{Creativity}                   & AI lacks creativity and originality               & 75.0\,\% & 8 \\ 
    \bottomrule
  \end{tabular}
\end{table}

Below, we elaborate each category, weaving insights and direct participant quotes from the retrospective and the research roadmap sessions of the workshop\cite{herda2025ai}:
\begin{itemize}
    \item \textbf{Tooling Challenges (F1)}: Participants described a “paradox of choice” in an ever-shifting landscape of proprietary copilots, chatbots and orchestration frameworks. The top grievance - “too many tools, unclear which to use” - captured nearly three quarters of all F1 votes. Root causes include fragmented vendor ecosystems and “capitalism baked into new tools,” which drive rapid, sometimes superficial differentiation. Frequent model updates further destabilize workflows, while mandated enterprise licenses (e.g., enforced use of MS Copilot) limit experimentation.
    \item \textbf{Governance \& Compliance (F2)}: More than half the voters flagged opaque data-handling policies and unclear regulatory boundaries as their chief worry. Practitioners “don’t know what’s behind the checkbox” when opting-out of model telemetry, amplifying concerns about GDPR, intellectual-property leakage and emerging EU AI Act obligations. The roadmap session highlighted the need for sandboxed evaluations of local models and agent-based audit trails to restore trust.
    \item \textbf{Team \& Process Misalignment (F3)}: Even when tools are available, AI often “doesn’t yield valuable outcomes.” Participants traced this to missing success criteria, over-reliance by junior developers and premature abandonment after early failures. Without explicit feedback loops, AI suggestions can ossify poor architectural choices or reinforce existing bottlenecks.
    \item \textbf{Data \& Model Quality Issues (F4)}: Hallucinations, stale context windows and inconsistent data schemas undermine confidence in AI output. Two-thirds of F4 votes targeted unreliable answers, while others lamented the difficulty of judging response quality in real time. Teams voiced a “fear of investing in multiple LLMs” without mechanisms to validate and cross-check results.
    \item \textbf{Knowledge \& Prompting Gaps (F5)}: The single most-voted sub-challenge across all categories was the lack of prompting skills (78.6 \%). Workshop stories described prompt crafting as “writing code in natural language,” yet few organizations provide systematic training. Shadow-agent concepts and role-specific onboarding emerged as near-term remedies during the deep-dive.
    \item \textbf{Creativity (F6)}: Finally, three-quarters of votes in F6 concerned AI’s tendency to regress toward the mean, producing “bland” or derivative ideas. Because foundation models are trained on historical artifacts, teams struggle to use them for divergent ideation or blue-sky design. Participants proposed human-AI pairing patterns and multi-modal toolchains to unlock co-creative potential.
\end{itemize}

\subsection{Implications for the Research Roadmap}

Collectively, these challenges expose a socio-technical gap: while AI tooling advances rapidly, organizational practices, regulatory guidance, and human skills lag. Each category of frustration reveals gaps that pose a direct risk to the integrity of agile principles, including transparency, adaptability, and value-driven delivery. To chart a path forward, we deliberately reframed these pain points into actionable research challenges, serving as a foundation for the co-creation of a research roadmap structured around five themes, 
each of which directly reflects one or more practitioner-identified frustration categories (F1–F6), ensuring that subsequent research actions address the pains most keenly felt by practitioners. Bridging this gap demands interdisciplinary effort, drawing upon software engineering foundations with insights from human–computer interaction (HCI), organizational psychology, and digital ethics - an agenda the remainder of this paper now develops.

\section{Roadmap for Human-Centered AI Integration in Agile Software Development}
To address the frustrations and challenges reported in the workshop \cite{herda2025ai}, we articulate a vision of agile software development where GenAI systems are not merely assistants, but integrated, context-aware teammates that actively contribute to team cognition, decision-making, and creativity. The roadmap promotes an agenda that emphasizes human-centeredness \cite{shneiderman2022human}, trustworthiness \cite{smuha2019eu,jobin2019global,lo2023trustworthy}, and sustained impact \cite{davenport2018artificial} as essential.  

First, we envision a transition from isolated smart tools to adaptive GenAI teammates. That is to say, systems are capable of understanding team workflows, learning from interactions, and contributing contextually relevant support throughout the agile lifecycle  \cite{xiao2024generative,richter2025there}. As emphasized by Lo et al. \cite{lo2023trustworthy}, synergistic AI requires deep integration with team cognition, historical artifacts, and evolving objectives.

Second, human–AI collaboration should preserve the psychological flow \cite{csikszentmihalyi1990flow} states that are critical to developer productivity and team coherence. Rather than interrupting or fragmenting work, GenAI should enhance team dynamics by anticipating needs, complementing skills, and enabling more focused execution \cite{noda2023devex}. This requires sociocognitive architectures \cite{gupta2025fostering} that respect human cognitive rhythms and promote trust through explainability and reliability to unlock the potential of human-machine intelligence.

Third, agile environments demand AI systems that can evolve alongside human practices. We therefore advocate for feedback loops \cite{mosqueira2023human} in which both human critique and machine self-assessment play continuous roles in refining AI behavior. Iterative feedback loops are essential to ensure that AI supports rather than constrains exploration, creativity, and adaptation \cite{adadi2018peeking}. 

These design principles form the conceptual foundation for the research roadmap. Building directly on the frustrations synthesized in Section 2 and guided by the vision articulated above, the roadmap is structured into five interrelated themes. 

\begin{itemize}
\item Theme 1: Tooling Ecosystem \& Integration
\item Theme 2: Human Factors, AI Literacy \& Team Mindset
\item Theme 3: Governance, Compliance \& Safe AI Use
\item Theme 4: Value Realization \& Evaluation
\item Theme 5: Creativity \& Multimodality in Agile AI
\end{itemize}

Each theme addresses a distinct set of socio-technical challenges, mapped to one or more frustration categories (F*). The themes integrate both the most highly rated challenges in Table~\ref{tab:frustration-categories} and other significant concerns identified by workshop participants \cite{herda2025ai}. Within each theme, we articulate a set of short-term research actions alongside longer-term research directions that point toward transformative shifts in integrating GenAI in agile practice. For clarity, while our primary focus is on GenAI, it is used interchangeably with AI in some contexts. Given the rapid advancement of AI technology, the research roadmap remains open to exploring different relevant AI tools. Together, these themes define a research agenda that is both responsive to practitioner needs and shaping the future of GenAI in agile practice.


\subsection{Roadmap theme 1: Tooling Ecosystem \& Integration}
Challenge 1 (F1): Tool overload and lack of clarity on which tools to use\\

The proliferation of GenAI tools presents agile teams with a paradox of choice \cite{schwartz2015paradox}. While new capabilities of LLMs are continuously introduced, the abundance of disconnected, overlapping, and non-interoperable tools have led to a fragmented ecosystem that imposes cognitive load. Teams often struggle with tool fatigue and constant context switching as they juggle multiple platforms. The fragmented toolchains force developers to toggle between many interfaces, which can disrupt workflow and the productivity. 
As observed in recent empirical studies \cite{li2024survey}\cite{jin2024integrating}, the lack of unified guidance on when and how to use AI tools in agile workflows hampers adoption and often leads to suboptimal or abandoned implementations. Participants in the XP2025 workshop identified "too many tools, unclear which to use" as the most-voted challenge (73.3\% of votes) in the category of frustration F1. This sentiment echoes a broader concern in the software engineering community that the rapid pace of AI tooling outstrips organizations’ ability to discern when and how to use these tools optimally.

To address these immediate needs, we propose 
short-term research actions focused on mapping the tool ecosystem and providing actionable guidance for tool adoption. These efforts will produce tangible resources to reduce decision paralysis among practitioners while also establishing an empirical baseline for how the AI tool landscape is evolving. The short-term research actions are elaborated as follows.

\begin{itemize}
    \item \textbf{Systematic review and taxonomy of AI tools in agile contexts}:
    This involves conducting a comprehensive review of both academic and industrial sources to identify and categorize GenAI tools being used across agile development activities. Recent work has shown that GenAI technologies are already being applied throughout the software lifecycle, from requirements and design to coding assistance, automated testing, and even agile retrospectives \cite{cornide2025generative}. Building on such insights, using mixed methods such as literature review, expert interviews, and tool feature analysis, this research will develop a structured taxonomy linking specific tool capabilities to agile practices, e.g., code generation tools for implementation, AI test generation for QA, planning assistants for sprint planning. The expected outcome is a publicly available reference that mitigates decision fatigue and supports more informed tool adoption and onboarding strategies within agile teams.
    
    \item \textbf{Open-access tool selection guide}:
    This action proposes the development of a practical decision-support guide to help agile teams select appropriate AI tools tailored to their context, such as task requirements, collaboration patterns, organizational constraints, and regulatory and privacy considerations. The guide will be informed by empirical heuristics derived from case studies and expert consultation and will be delivered as an interactive resource, such as a living document or web-based decision aid. By offering just-in-time guidance tailored to user needs, the guide aims to alleviate tool fatigue and enhance confidence in AI adoption across varying Agile maturity levels.
\end{itemize}

While these short-term initiatives address the need for clarity in a fragmented tool landscape, a complementary long-term priority lies in shifting educational focus from tool-specific training to fundamental AI capabilities. Rather than emphasizing mastery of specific platforms that may become obsolete, organizations should cultivate foundational AI literacy that transcends individual tools. This approach enables practitioners to adapt seamlessly to any AI environment—from cutting-edge platforms to constrained corporate systems with limited tool availability. By developing competency in core AI concepts, prompt engineering, and integration principles, agile teams build transferable skills that remain valuable regardless of technological shifts. This capability-centered education strategy not only reduces the cognitive burden of constant tool switching but also transforms the paradox of choice from a barrier into an opportunity for flexible adaptation.

Looking toward the future, the long-term research direction focuses on more integrated, adaptive, and intelligent approaches to tooling in agile environments. These research directions aim to explore how teams interact with AI across multiple workflows, promoting both scalability and sustainability. In addition, successful AI adoption depends on the maturity of underlying orchestration platforms and a clear understanding of associated trade-offs such as latency, reliability, and the “butterfly effect” in multi-agent settings where small variations in one agent’s response can cascade through others, influencing outcomes in unpredictable ways. Therefore, the proposed long-term research directions are as follows.

\begin{itemize}
    \item \textbf{Multi-agent model selection interface}: This research aims to prototype a unified user interface that intelligently selects the most suitable AI model or agent for a given agile task, based on user intent, project history, and contextual signals. Leveraging orchestration frameworks such as LangChain \cite{Langchain} or Haystack \cite{Haystack} and integrating routing logic that is sensitive to task type and team preferences, this interface builds on advancements in multi-agent systems including SWE-Agent \cite{yang2024swe} and OpenDevin \cite{wang2024opendevin}. The envisioned result is a reduction in cognitive load for team members and an improvement in delegation efficiency when using AI across iterative development cycles.
    \item \textbf{Longitudinal evaluation of integrated toolchains}: This action involves conducting longitudinal field studies in agile organizations to assess the impact of integrated AI toolchains over multiple sprints. Using quantitative metrics, e.g., code quality, velocity, rework rate, and qualitative indicators, e.g., developer satisfaction, trust in AI, the studies will evaluate how AI tools influence coordination, productivity, and technical debt. Drawing upon prior work \cite{niederman2016co,yalccin2023studying} on developer experience and tool co-evolution, the goal is to provide evidence-based insights that guide sustainable and value-driven AI tool integration. Notably, a systematic review \cite{cornide2025generative} highlighted a lack of empirical evaluations on GenAI’s long-term impact in software teams, calling for exactly this kind of study.
    \item \textbf{Domain-specific adaptation of AI models}: This research investigates methods for tailoring AI models to the requirements of domain-specific agile practices, such as those in finance, healthcare, or logistics. Approaches may include fine-tuning models with domain-relevant corpora, applying retrieval-augmented generation (RAG) techniques, and embedding organizational knowledge using protocols such as the model context protocol (MCP) into prompt design. The expected benefit is enhanced relevance, accuracy, and contextual fidelity of AI-generated artifacts, contributing to better alignment with real-world constraints and expectations in regulated or specialized agile environments.
\end{itemize}

Table \ref{tab:tooling ecosystem} summarizes the short-term and long-term research activities.

\begin{table}[ht]
\centering
\caption{Tooling Ecosystem \& Integration}
\label{tab:tooling ecosystem}
\begin{tabularx}{\linewidth}{|l|X|X|>{\centering\arraybackslash}p{2cm}|}
\hline
\textbf{Timeframe} & \textbf{Research} & \textbf{Expected Outcomes} & \textbf{Challenges addressed} \\
\hline
\multirow{2}{*}{Short-term actions}
  & Conduct a systematic review of existing AI tools relevant to agile practices
  & Tools taxonomies and their mapping to agile activities
  & 1 \\
\cline{2-4}
  & Develop an open-access guide mapping tools for specific agile activities
  & Shared resources to reduce tool fatigue and misalignment
  & 1 \\
\hline
\multirow{3}{*}{Long-term  directions}
  & Prototype a UI layer that dynamically selects the best AI agent for a task (multi-agent model picker)
  & Increased adoption and trust in AI tools
  & 1 \\
\cline{2-4}
  & Evaluate integrated toolchains through longitudinal field studies in agile teams
  & Empirical evidence on productivity impact
  & 1 \\
\cline{2-4}
  & Explore methods to augment domain-specific knowledge in models
  & Domain-specific models
  & 1 \\
\hline
\end{tabularx}
\end{table}

\subsection{Roadmap theme 2: Human Factors, AI Literacy \& Team Mindset}
Challenge 2 (F5): Teams lack prompting skills and understanding of AI's strengths/limitations \\
Challenge 3 (F3): Poor integration into team workflows\\

The successful integration of GenAI into agile environments depends not only on technical maturity but also on the human readiness of development teams. Foundational AI literacy \cite{pan2025ai}, as discussed in Roadmap Theme 1, provides a necessary baseline. However, effective collaboration with GenAI systems also requires cultivating shared mental models and an adaptive team mindset \cite{andrews2023role}. 

Despite the increasing availability of AI tools, many teams struggle to derive value from them due to unclear expectations, inconsistent understanding of system capabilities and limitations\cite{andrews2023role,desolda2025understanding,bahi2024integrating}, and poor alignment between AI functionality and agile workflows. The lack of prompting skills and uncertainty about appropriate use cases were frequently cited in the workshop as barriers to experimentation and trust. These limitations not only hinder adoption but can also lead to misuse or superficial engagement with GenAI tools \cite{pan2025ai}. 

Bridging this socio-technical gap demands more than technical training. It requires building a team-wide capacity for responsible, reflective, and context-aware AI use. To this end, we propose a series of short-term research actions aimed at enhancing team-level AI literacy, shared understanding, and supporting the responsible use of AI by agile teams. The goal is to equip project teams with the skills and mindsets needed to interact
effectively with AI systems across varying development contexts.

\begin{itemize}
    \item \textbf{Prompting skills assessment and role-specific training interventions}: This research action focuses on identifying current prompting competencies within agile teams and developing tailored educational interventions for different roles such as developers, product owners, and Scrum masters. Using surveys, task-based assessments, and observational studies, researchers can benchmark the effectiveness of prompting strategies and uncover gaps in understanding. The resulting data will inform the design of context-specific training modules that include hands-on exercises and reflective practices, thereby supporting the cultivation of robust human–AI collaboration habits across the agile project lifecycle.
    
    \item \textbf{Designing onboarding frameworks and team-level AI literacy workshops}: This initiative centers on the development and evaluation of onboarding practices that introduce AI capabilities to new team members in a structured and responsible manner. It also involves organizing team-based workshops that build shared vocabulary, clarify model limitations, and address common misconceptions. Drawing on research in team cognition and socio-technical systems, these workshops are intended to foster collective literacy and trust, enabling teams to co-evolve with AI rather than treat it as a static external tool \cite{andrews2023role}.
\end{itemize}

For more transformative change over time, long-term research directions should explore the integration of AI as a reflective collaborator within agile teams and examine how team mindsets shift with continued use. These directions aim to study the dynamic interplay between evolving tools, human learning, and organizational adaptation.

\begin{itemize}
    \item \textbf{Development and deployment of "shadow AI" agents in agile Teams}: This line of research envisions AI agents that operate in a non-intrusive, parallel mode in project team activities such as sprint planning or retrospectives. These "shadow agents" observe decision-making processes and offer post-hoc feedback, alternative suggestions, or strategic prompts based on team goals and prior behaviors. The objective is to explore how such agents can support reflective practices, improve decision quality, and encourage more deliberate and explainable use of AI in complex team settings.
    \item \textbf{Empirical studies on organizational change and mindset evolution}: This long-term research investigates how team culture, norms, and mindset evolve as AI becomes integrated into agile projects. Longitudinal studies, which involve interviews, ethnography, and organizational surveys, can shed light on how AI adoption affects collaboration dynamics, leadership expectations, and notions of authorship and accountability. The findings will inform the design of change management strategies and leadership interventions tailored to AI-integrated agile organizations.
    
    \item \textbf{Human-AI partnership models for agile roles}: This direction proposes the co-design and empirical validation of partnership frameworks that articulate complementary strengths between humans and AI across agile roles. The introduction of GenAI poses questions about team structure, e.g. How do we conceptualize an AI “teammate”? In what tasks should AI take the lead, and where must a human always intervene? To address such questions, the research can be conducted to develop partnership models that define clear allocations of decision authority, quality checking, and knowledge sharing between humans and AI. By defining such models and human-AI interaction, teams can avoid both over-reliance on AI and under-utilization of its capabilities. Crucially, these models should be evaluated in practice to refine how trust is calibrated. The aim is to shift the framing of AI from a tool to a teammate, while still preserving critical human oversight and ethical responsibility.

\end{itemize}

A summary of the short-term and long-term research activities is presented in Table \ref{tab:human factors}. 

\begin{table}[ht]
  \centering
  \caption{Human Factors, AI Literacy \& Team Mindset}
  \label{tab:human factors}
\begin{tabularx}{\linewidth}{|l|X|X|>{\centering\arraybackslash}p{2cm}|}
    \hline
    \textbf{Timeframe} & \textbf{Research} & \textbf{Expected Outcomes} & \textbf{Challenges addressed} \\
    \hline
    \multirow{2}{*}{Short-term actions}
      & Conduct workshops, surveys, walkthroughs, etc.\ to evaluate current prompting practices and AI literacy levels in agile teams
      & Baseline understanding of skill gaps
      & 2 \\
    \cline{2-4}
      & Design and deploy team-focused AI training and onboarding modules
      & AI literacy programs
      & 2 \\
    \hline
    \multirow{3}{*}{Long-term directions}
      & Introduce “shadow agents” in team projects as silent observers to give feedback and recommendations
      & Experience with AI-supported collaboration
      & 3 \\
    \cline{2-4}
      & Research on team organizational mindsets and cultural readiness for AI adoption
      & Best practice on change management
      & 2, 3 \\
    \cline{2-4}
      & Research on empirical evaluation of partnership between human and AI
      & AI shifted from a tool to a teammate
      & 2, 3 \\

    \hline
  \end{tabularx}
\end{table}

\subsection{Roadmap theme 3: Governance, Compliance \& Safe AI Use}
Challenge 4 (F2): Lack of clarity in data protection and legal compliance\\
Challenge 5 (F4): Lack of trust in tool\\

Along with increasingly integrating GenAI in agile workflows, concerns about legal, ethical, and procedural governance emerge as central challenges. Teams must navigate a complex landscape of organizational policies, industry standards, and evolving regulatory frameworks such as the EU AI Act \cite{act2024eu}, ISO/IEC 42001 on AI management systems \cite{benraouane2024ai}, and the IEEE 7000 series on ethical system design \cite{spiekermann2021expect}. While these frameworks emphasize accountability, explainability, and human oversight, most agile teams lack clear guidelines on how to operationalize them in day-to-day development. Teams face growing uncertainty about how to ensure compliance with data protection regulations, how to manage intellectual property in AI-generated artifacts, and how to maintain transparency in AI-driven decisions. Unregulated AI would pose a threat, as it would be impossible to effectively monitor its effects on the economy, society, and environment \cite{zhao2023artificial}. 

Concrete legal and procedural challenges include the need to comply with the General Data Protection Regulation (GDPR) in the European Union, which governs personal data processing and imposes strict obligations for transparency and accountability \cite{voigt2017eu}. AI systems that process user stories or meeting transcripts must implement privacy-by-design principles \cite{voigt2017eu} to ensure lawful data use.

From an intellectual property perspective, agile teams require clarity on the ownership of AI-generated artifacts, such as user stories, test cases, or code suggestions, particularly when these contents are derived from models trained on public or proprietary datasets. Recent debates around the use of copyrighted training data and the rights to AI-generated outputs emphasize the urgency for legal guidance \cite{samuelson2024fair}.

Ethically, there is an increasing demand for AI models to be explainable and auditable to align with principles of fairness, accountability, and transparency (FAT) \cite{goodman2017european,adadi2018peeking,gunning2019xai}. This includes making clear how recommendations are generated and allowing stakeholders to challenge automated decisions. Governance frameworks such as the OECD AI Principles and the EU AI Act highlight the importance of risk classification, human oversight, and post-deployment monitoring to ensure the responsible use of AI in dynamic environments like agile software teams.

Participants in the workshop identified unclear privacy boundaries and lack of transparency as critical barriers to the trustworthy use of AI. These concerns not only slow down adoption but can also result in unintentional policy violations or reputational risks. There is a clear need to establish frameworks, guidelines, and practical tools that enable teams to responsibly deploy AI while upholding ethical standards and regulatory compliance. 

Notably, we also observe the emergence of new human roles within agile product teams that focus on navigating legal and regulatory complexities related to AI. These roles include AI compliance officers, data governance leads, and ethics champions. They are increasingly embedded in cross-functional teams to ensure proactive alignment with evolving policy and legal expectations. Their presence reflects a shift toward organizationally grounded compliance practices that complement technical safeguards and foster shared responsibility for trustworthy AI use.

To mitigate immediate risks and provide actionable guidance to teams, we propose a set of short-term research efforts aimed at clarifying governance expectations and creating controlled environments where AI use can be evaluated safely.

\begin{itemize}
    \item \textbf{Sandbox environments for AI evaluation}: This research action involves the development of isolated testing environments where teams can experiment with AI tools on synthetic or anonymized datasets to evaluate their behavior and outputs in a risk-free setting. These sandboxes can help identify potential compliance breaches or unintended consequences prior to real-world deployment. By designing controlled experiments within these environments, researchers and practitioners can collaboratively assess data handling practices, logging transparency, and model interpretability. The resulting insights will enable the development of usage guidelines tailored to Agile contexts and aligned with privacy legislation such as GDPR.

    \item \textbf{Practitioner-focused briefs on AI regulation and intellectual property}: This initiative seeks to create concise, accessible briefs for agile practitioners summarizing the implications of current regulations related to AI, including privacy laws, data residency requirements, and intellectual property rights. Drawing from legal scholarship and case studies in software engineering, these briefs would offer context-specific scenarios and recommended practices for compliant AI adoption. By improving awareness and accessibility of legal knowledge, this action aims to close the gap between policy and practice in fast-paced agile environments. These resources would also support the work of newly emerging legal and ethical roles within teams, strengthening capacity for responsible innovation.
    \item \textbf{Emergence of new governance roles in agile teams}: This research action focuses on defining and supporting the roles, such as AI compliance specialists and ethics leads, that are becoming critical in agile teams. This involves studying their responsibilities, organizational integration patterns, and the ways they can enhance legal agility within the team. The goal is to develop guidance on how these roles can effectively facilitate responsible AI adoption and ensure proactive alignment with evolving policy and legal expectations.
\end{itemize}

In the long-term, research should address structural governance mechanisms that support adaptive and transparent AI oversight. These directions aim to embed trustworthy practices at the architectural and organizational level.

\begin{itemize}
    \item \textbf{Transparent AI audit mechanisms within agile decision-support tools}: This research direction focuses on designing and testing mechanisms that enable clear traceability and auditability of AI decisions within agile software development tools. Drawing on principles from explainable AI and software compliance, the goal is to develop functionalities that support post-hoc reviews of AI recommendations and detect potential policy violations. This ensures AI systems remain accountable and verifiable across varying contexts and regulatory landscapes. 

    \item \textbf{Explore agent-based governance frameworks}: This research direction envisions the development of extensible governance frameworks that combine logging and auditability with automated policy enforcement through agent-based infrastructure. These frameworks would support real-time detection of policy violations and provide continuous monitoring of model usage within Agile workflows. The integration of monitoring agents capable of flagging compliance risks and guiding responsible usage exemplifies how governance can be embedded at the infrastructure level without overwhelming human intervention. 
\end{itemize}

A summary of the short-term and long-term research activities is presented in Table \ref{tab:governance-overview}.

\begin{table}[H]
  \centering
  \caption{Governance, Compliance \& Safe AI Use}
  \label{tab:governance-overview}
\begin{tabularx}{\linewidth}{|l|X|X|>{\centering\arraybackslash}p{2cm}|}
    \hline
    \textbf{Timeframe} & \textbf{Research} & \textbf{Expected Outcomes} & \textbf{Challenges addressed} \\
    \hline
    \multirow{3}{*}{Short-term actions}
      & Develop sandbox environments for experimentation with local or lightweight AI models using synthetic data 
      & Safe, risk-free testing environment, improved team awareness of data flows and compliance risks
      & 4 \\
    \cline{2-4}
      & Develop practitioner briefs summarizing legal, privacy, and IP considerations for AI in agile contexts 
      & Accessible guidance on GDPR, IP ownership, and auditability tailored to agile teams 
      & 4 \\
    \cline{2-4}
      & Identify and support emerging human roles for AI governance in agile teams (e.g., AI compliance officers, ethics leads)
      & Role definitions, organizational patterns, increased legal agility and team accountability
      & 4, 5 \\
    \hline
    \multirow{2}{*}{Long-term directions}
      & Design and test transparent AI audit mechanisms within agile decision-support tools
      & Improved traceability and compliance support, increased trust in AI-supported workflows
      & 4, 5 \\
    \cline{2-4}
      & Explore agent-based governance frameworks with customizable oversight capabilities
      & Autonomous, auditable AI deployment
      & 4, 5 \\
    \hline
  \end{tabularx}
\end{table}

\subsection{Roadmap theme 4: Value Realization \& Evaluation}
Challenge 6 (F3): Difficulty in quantifying AI's real business value\\
Challenge 7 (F3): Lack of clear success criteria in agile AI adoption\\

Despite the enthusiasm surrounding AI adoption in agile contexts, many teams struggle to demonstrate clear, measurable value from their AI investments \cite{perifanis2023investigating,pandey2021roi}. The workshop participants identified the need for better methods to quantify AI impact, particularly in terms of business value, team velocity, and long-term maintainability. Current success metrics are often too coarse, short-sighted, or misaligned with actual team goals, leading to premature abandonment or unjustified reliance on AI tools. Furthermore, teams lack systematic approaches to monitor, evaluate, and refine AI-supported processes over time. A dedicated research focus on value realization is critical to ensuring that AI contributes meaningfully and sustainably to agile practices. To address these issues in the short term, we propose practical, evaluative strategies that allow agile teams to articulate what value means in their context and gather empirical data to support informed decision-making.

\begin{itemize}
    \item \textbf{Contextual definition of success criteria for AI in agile}: This research action focuses on helping teams define success criteria for AI adoption that are tailored to their project goals, maturity level, and organizational culture. Drawing from goal-oriented requirements engineering \cite{van2001goal}, researchers can collaborate with teams to co-create value metrics that balance qualitative insights with quantitative indicators. These criteria should span not only productivity and quality, but also technical debt trends, developer satisfaction, design decision latency, and team turnover risk. The resulting frameworks will allow teams to align expectations, monitor progress, and make better-informed decisions about scaling or pivoting their AI usage.

    \item \textbf{Multi-faceted evaluation frameworks for AI-enabled agile practices}: This initiative aims to develop and validate structured frameworks that allow for continuous assessment of AI contributions across agile activities such as backlog refinement, test generation, and sprint retrospectives. The frameworks will combine multiple evaluation methods, including longitudinal metrics, developer diaries, and stakeholder interviews, to capture both immediate and cumulative impacts. Emphasis will be placed on interpretability and transparency so that results can inform actionable improvements. These frameworks will offer a foundation for organizational learning and evidence-based scaling of AI across projects.
\end{itemize}

Over the longer term, the research agenda must extend toward rigorous and adaptive methods for value modeling and impact estimation, enabling organizations to assess the return on investment and broader implications of AI integration.

\begin{itemize}
    \item \textbf{AI-driven value tracking and feedback loops}: This research proposes the development of systems and models that enable dynamic tracking of AI’s contribution to value creation over time. Building on techniques from continuous software engineering and DevOps observability, the approach involves integrating AI observability tools with agile analytics dashboards. These systems would monitor aspects such as productivity, quality, maintainability, technical debt trajectory, decision latency, and staff retention patterns. 
    Incorporating feedback loops, these insights could be used to adapt AI usage, recalibrate team practices, or refine tool configurations. Ultimately, the goal is to create a robust empirical foundation for understanding when, how, and why AI delivers value in agile contexts.
    \item \textbf{Economic and organizational models of AI-Agile synergy}: This direction explores how economic theories and organizational science can be applied to model the costs, risks, and benefits of AI integration in agile ecosystems. Using case study research, simulation models, and econometric analyses, researchers can investigate how AI affects productivity dynamics, team coordination, innovation cycles, and risk management. These models can guide strategic decision-making at both the team and enterprise levels by offering predictive insights about return on investment and long-term sustainability. The outcome is a richer understanding of AI effects and a more nuanced framework for evaluating its role in future software engineering practices.
\end{itemize}

Table \ref{tab:value realization} summarizes the research activities on theme of value realization and evaluation.

\begin{table}[ht]
  \centering
  \caption{Value Realization \& Evaluation}
  \label{tab:value realization}
\begin{tabularx}{\linewidth}{|l|X|X|>{\centering\arraybackslash}p{2cm}|}
    \hline
    \textbf{Timeframe} & \textbf{Research} & \textbf{Expected Outcomes} & \textbf{Challenges addressed} \\
    \hline
    \multirow{2}{*}{Short-term actions}
      & Define success criteria for AI adoption with agile teams based on project context and team maturity
      & Improved clarity on AI value expectations; tailored success metrics
      & 6 \\
    \cline{2-4}
      & Develop multi-faceted evaluation frameworks  
      & Evaluation toolkit for assessing short-term AI impact
      & 6, 7 \\
    \hline
    \multirow{2}{*}{Long-term directions}
      & Build feedback loops and analytics dashboards that quantify AI impact on team productivity and decision quality
      & Better decision support for AI integration
      & 7 \\
    \cline{2-4}
      & Explore economic and organizational models to evaluate costs, benefits, and long-term impacts of AI integration in agile ecosystems 
      & Predictive insights into long-term value, sustainability, and strategic decision-making for AI adoption
      & 6, 7 \\
    \hline
  \end{tabularx}
\end{table}

\subsection{Roadmap theme 5: Creativity \& Multimodality in Agile AI}

Challenge 8 (F6): AI is constrained by training data\\
Challenge 9 (F3): Teams don’t exploit creative potential in ideation phases\\

AI systems often reflect and reinforce existing patterns in their training data, which can limit their capacity to support exploratory, generative, and divergent thinking. At the same time, agile teams frequently underutilize the creative potential of AI tools during early-stage ideation and design. Participants in the XP2025 workshop emphasized the need to expand the use of AI beyond routine or efficiency-oriented tasks, and to explore how AI can act as a catalyst for creativity, innovation, and multimodal problem-solving. Embracing this potential requires both methodological experimentation and reimagining how AI contributes to human-centered design processes. To build foundational understanding and stimulate new applications, short-term research efforts should investigate practical use cases and tool combinations that enable creativity and design ideation.

\begin{itemize}
    \item \textbf{Exploratory case studies on AI-augmented design and ideation}: This research direction involves conducting case studies with agile teams experimenting with AI tools to support early-stage ideation, user story development, and product concept visualization. The goal is to document effective practices, identify common challenges, and synthesize insights into a catalog of best practices. These studies would also explore how team roles and dynamics evolve when AI tools are used to generate, refine, or critique design alternatives.

    \item \textbf{Experimental on multimodal AI in design activities}: This research focuses on evaluating the potential of combining text-based, visual, and code-generating AI models in collaborative design scenarios. Experimental setups involving controlled tasks can help assess how multimodal tools (e.g., image generation, diagram synthesis, prototyping assistants) influence creativity, collaboration, and design diversity. The outcomes will help identify effective toolchains and use cases that extend the current boundaries of AI support in agile design processes.
\end{itemize}

In the long term, research must examine how co-creative relationships between AI and agile teams evolve and what structures best support sustained innovation.

\begin{itemize}
    \item \textbf{Designing and Evaluating Co-Creative Workflows for Agile Teams}: This research proposes the development of collaborative workflows in which AI tools serve as creative partners throughout the design process. These workflows may include turn-based ideation, AI-generated provocations, or creative retrospectives enhanced by generative models. The goal is to formalize and evaluate interaction patterns where human agency is preserved while benefiting from AI's associative and recombinatory capabilities. Empirical studies can assess whether such workflows improve team innovation, satisfaction, and inclusivity.

    \item \textbf{Longitudinal studies on creativity and innovation in AI-augmented teams}: This research direction aims to understand the sustained impact of AI on creative thinking and innovation outcomes in agile teams. Using longitudinal case studies and qualitative tracking methods, researchers can explore how team perceptions, processes, and outputs change over time when AI is embedded in ideation, design sprints, or product strategy discussions. The insights will inform the development of guidelines and frameworks for integrating creativity as a core principle in human-AI collaboration.

\end{itemize}

Table \ref{tab:creative-overview} summarizes the research activities on the theme of creativity \& multimodality in Agile AI.

\begin{table}[ht]
  \centering
  \caption{Creativity \& Multimodality in Agile AI}
  \label{tab:creative-overview}
 \begin{tabularx}{\linewidth}{|l|X|X|>{\centering\arraybackslash}p{2cm}|}
    \hline
    \textbf{Timeframe} & \textbf{Research} & \textbf{Expected Outcomes} & \textbf{Challenges addressed} \\
    \hline
    \multirow{2}{*}{Short-term actions}
      & Case study on AI supporting creative ideation and design activities in agile
      & Best practices catalog
      & 8, 9 \\
    \cline{2-4}
      & Run experiments with multimodal AI tools—combining text, image, and code—for early-stage design
      & Use cases for enhanced ideation
      & 9 \\
    \hline
    \multirow{2}{*}{Long-term directions}
      & Develop and promote co-creative workflows where AI actively collaborates in the design
      & New models of collaborative creativity
      & 9 \\
    \cline{2-4}
      & Evaluate long-term creative outcomes and innovation potential in AI-supported agile teams
      & Guidelines for Agile + Creativity
      & 9 \\
    \hline
    \end{tabularx}
\end{table}

\section{Implementation Enablers}

The successful realization of this research roadmap hinges on several foundational enablers that support experimentation, collaboration, and evaluation. First, there is a pressing need to establish dedicated AI for agile software development (AI4Agile) research testbeds. These environments should simulate agile project settings and offer support for controlled experiments with AI tools, enabling rigorous testing of hypotheses and replicability of results. Such testbeds will facilitate comparative evaluation across different contexts, tools, and team configurations.

Second, progress in integrating AI into agile projects depends on the availability of high-quality, annotated datasets. These datasets should include diverse agile artifacts such as user stories, commit messages, meeting transcripts, and design prototypes, all of which are essential for training and benchmarking AI models. Open and well-documented datasets will also encourage reuse and reproducibility in academic and industrial research.

Third, there is a need for robust evaluation frameworks that can capture the multifaceted impacts of AI tools in agile environments. These frameworks should include both qualitative and quantitative metrics, covering aspects such as productivity, team collaboration, trust, fairness, and ethical alignment. Methodologies should enable longitudinal tracking and cross-team comparisons to inform broader conclusions about AI's value and limitations.

Finally, collaboration must be enabled through shared infrastructure and open-source platforms. Repositories of tools, benchmark suites, prompts, and case studies will foster collective learning and accelerate adoption. By lowering the entry barrier to experimentation and promoting transparency, such infrastructure can amplify the community’s ability to iterate and refine best practices for human-centered AI in agile projects.

\section{Conclusion and call for collaboration}

This research roadmap reflects the collective insights of practitioners and researchers at the XP2025 workshop. We present an actionable research agenda for AI adoption within agile software development. This agenda consists of five interconnected themes, each addressing identified challenges with both short-term actions and long-term research directions. From integrating fragmented tools to establishing robust governance and fostering human-AI collaboration. These themes collectively outline how AI and agile practices intersect and co-evolve.

A key message of the roadmap is the necessity of an interdisciplinary approach. Researchers from software engineering, human-computer interaction, ethics, law, and organizational psychology must work closely with industry practitioners to develop solutions that are not only technically robust but also ethically sound and practically viable. These cross-sector partnerships are essential to resolve open issues related to explainability, privacy, trustworthiness, and support in complex, real-world development contexts.

While the workshop revealed practitioner concerns and guided future research, we also acknowledge the methodological limitations of this study. While the workshop involved participants with diverse professional and geographical backgrounds, its size and duration constrained the perspectives captured. Consequently, some critical perspectives, specifically from roles such as end-users, product managers, or compliance specialists, may have been underrepresented. Additionally, the process of translating workshop outputs into roadmap themes inherently involves a degree of interpretation by the authors of the study. To validate and expand this agenda, we should include empirical follow-ups in future work. Conducting surveys, interviews, and longitudinal case studies will be crucial to triangulate practitioner concerns and proactively identify new needs as GenAI technologies continue to evolve.

Realizing the potential of this roadmap requires collective action. We call on the research and practitioner community to contribute to shared benchmarks, open datasets, collaborative platforms, and rigorous longitudinal evaluations. Such initiatives are crucial for systematically assessing AI tools in real-world agile projects, tracking their long-term impacts, and ensuring responsible experimentation.

We specifically invite researchers, practitioners, tool developers, policymakers, and educators to jointly shape the future of adopting AI for agile software development. Multi-disciplinary partnerships are essential to ensure that AI innovations are guided by ethical principles, grounded in empirical evidence, and reflecting the real needs of agile teams. We especially encourage contributions that

\begin{itemize}
    \item Contribute to shared datasets, benchmark suites, or validation protocols.
    \item Test roadmap themes in industrial settings and report findings.
    \item Explore intersections with adjacent areas such as HCI, organizational behavior, and digital ethics.
    \item Strengthen capacity building by integrating AI topics into agile education and training.
\end{itemize}

Through these collaborative efforts, the roadmap can evolve into a living, evidence-based resource that both guides research and supports practitioners. Ultimately, our shared goal is to ensure that AI adoption reinforces the spirit of agility to shape the future of software engineering in a sustainable and human-centered way.
\subsubsection{}
\textbf{Acknowledgements.} \textit{We thank the participants and programme committee members of the XP2025 workshop for their valuable insights for developing this research roadmap. The authors also acknowledge the use of GenAI tools, including the OpenAI GPT-5 and Google Gemini 2.5 Pro model, to polish and refine the language of this manuscript. All research ideas, analyses, and interpretations remain solely the responsibility of the authors.}


\bibliographystyle{splncs}
\bibliography{references}

\end{document}